\documentclass[aps,prb,twocolumn,groupedaddress,showpacs]{revtex4}

\begin{document}
%\twocolumn[ \hsize\textwidth\columnwidth\hsize\csname
%@twocolumnfalse\endcsname
\title{Double-well potentials in current qubits}
\author{Mun Dae Kim, Dongkwan Shin, and Jongbae Hong}
\affiliation{School of Physics, Seoul National University, Seoul 151-747, Korea}
\date{\today}
\begin{abstract}
The effective potentials of the rf-SQUID and three-Josephson junction loop
with a penetrating external magnetic flux are studied.
Using the periodic boundary condition for the phase evolution of
the wave function of Cooper pairs, we obtain  new periodic
potentials with cusp barriers in contrast with the usual smooth double-well
potential.
The tunneling through the cusp barrier becomes dominant for a parameter regime
where the self inductance of the superconducting loop and the Josephson coupling
energy are large.
Calculating the tunneling frequency we see that
the cusp potential may offer a more efficient qubit.
\vspace{2mm}
\end{abstract}

\draft
\maketitle

\section{Introduction}

The rf-SQUID has attracted great attention  in connection with the
study of quantum two-level systems \cite{Leggett}. In particular it
has offered prospects for  macroscopic quantum phenomena,
such as quantum tunneling  and quantum coherence between two
quantum states \cite{Caldeira}. A frequently appearing structure is
the well-known double-well potential with a smooth barrier in some
parameter regime. Surprisingly, however, the inspection into the
rf-SQUID with precise boundary conditions has not been achieved.

A periodic boundary condition can be obtained by a gauge
invariance and the phase evolution by the wave vector of Cooper
pairs which is identical with the fluxoid quantization condition.
Using this boundary condition, we can obtain the effective
potential of an rf-SQUID and three-Josephson junction loop
which consist of the energy of moving
Cooper pairs and the Josephson coupling energy. The external flux
induces a persistent current of Cooper pairs in superconducting loop.
When the self inductance of the superconducting loop
and the Josephson coupling energy are large, the total flux,
sum of the external flux and induced flux, become quantized,
while the fluxoid quantization condition is always satisfied.

Considering the energy of  Cooper pairs, the effective
potential of the system shows a new periodic structure consisting
of  cusp and smooth barriers. In one extreme, the tunneling through
smooth-barrier potential is dominant, which is the usual case
\cite{Leggett,Caldeira,Kurk,Friedman,Mooij,Wal}, while, in the other
extreme, the cusp barrier potential is more appropriate in
tunneling phenomena. Both may  be used as a qubit in quantum
computing \cite{Nielsen,Makhlin}. We compare the tunneling
rates for the usual smooth-barrier potential and the cusp-barrier
potential and show that the latter may offer more effective qubit.

\section{rf-SQUID}

\begin{figure}[b]
\vspace*{7cm} \includegraphics{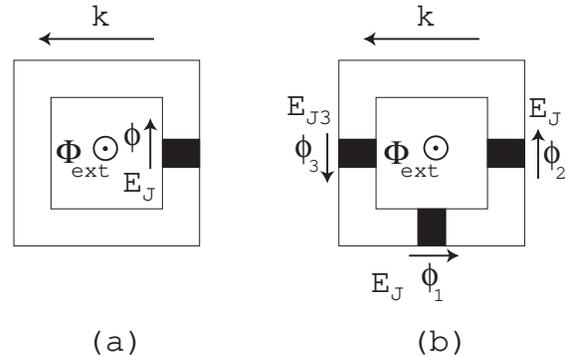}
\vspace*{-2cm} \caption{(a)
rf-SQUID and (b)  three-Josephson junction loop with the Josephson
coupling energy $E_J$'s and the wave vector of Cooper pairs $k$.}
\label{fig:SingQubit}
\end{figure}

We first consider an rf-SQUID shown in Fig. \ref{fig:SingQubit} (a).
We derive the effective potential of the system showing
double-well shapes with both smooth and cusp barriers.
The current through the Josephson junction in Fig. \ref{fig:SingQubit} (a)
is written as
\begin{eqnarray}
\label{I}
I=-I_c\sin\phi+\frac{V}{R}+C\dot{V},
\end{eqnarray}
where $\phi$ is the superconducting phase
difference across the Josephson junction and
$I_c$, $R$, and $C$  are critical current, resistance, and
capacitance, respectively.
Using the relation, $I=(\Phi_t-\Phi_{\rm ext})/L_s$, and  the Josephson
voltage-phase relation, $V=-(\Phi_0/2\pi)\dot{\phi}$, Eq. (\ref{I})
becomes
\begin{equation}
\label{Phi}
\Phi_t=\Phi_{\rm ext}-L_sI_c\sin\phi-L_s\frac{\Phi_0}{2\pi}
\left(\frac{\dot{\phi}}{R}+C\ddot{\phi}\right).
\end{equation}
Here $\Phi_t$ is the total flux,  $\Phi_{\rm ext}$ the external flux,
$\Phi_{\rm ind}=L_sI$ the induced flux, $L_s$  the  self inductance of
the superconducting loop, and $\Phi_0=h/2e$  the unit flux quantum. When
$L_sI_c/\Phi_0<<1$, one can use the approximation $\phi\approx
2\pi\Phi_t/\Phi_0$ in Eq. (\ref{Phi}), which results in the usual
smooth double-well potential \cite{Kurk}.

%%%%%%%%%%%%%%%%%%%%%%%%%%%%%%%%%%%%%%%%%%%%%%%%%%%%%%%%%%%%%%%%%%%%%%%%%%%%%%%%%%%%%%%%%%%
%Figure 1
%%%%%%%%%%%%%%%%%%%%%%%%%%%%%%%%%%%%%%%%%%%%%%%%%%%%%%%%%%%%%%%%%%%%%%%%%%%%%%%%%%%%%%%%%%%

The fluxoid quantization  condition for  a thin superconducting loop
\cite{Tinkham},
\begin{equation}
\label{fluxoid}
-\Phi_t+(m_c/q_c)\oint \vec{v}_c \cdot d\vec{l}=n\Phi_0
\end{equation}
with $q_c=2e$, $m_c=2m_e$,
and the average velocity of Cooper pairs $\vec{v}_c$, can be represented
as the periodic boundary condition,
\begin{equation}
k_n{\it l}=2\pi n+2\pi f_t,
\end{equation}
determining the wave vector at a quantum state $n$,
where $k_n$ is the wave vector of the Cooper pairs, $l$  the
circumference of the loop, and $f_t\equiv \Phi_t/\Phi_0$.
Here, $k_nl$ is the phase accumulated along the circumference of the loop  \cite{Matveev}.
For a rf-SQUID in Fig. \ref{fig:SingQubit} (a) the boundary condition
with the  phase difference across the Josephson junction $\phi$ becomes
\begin{eqnarray}
\label{pbc} k_n{\it l}=2\pi n+2\pi f_t-\phi .
\end{eqnarray}
The current, on the other hand, is given by $I=-(n_cAq_c/m_c)\hbar
k_n$, where $n_c$ is the Cooper pair density, $A$ the cross
section,  and the induced flux $\Phi_{\rm ind}=L_sI$ becomes
\begin{eqnarray}
\label{ft}
\frac{\Phi_{\rm ind}}{\Phi_0}=-\frac{\gamma l}{2\pi}k_n,
\end{eqnarray}
where  $\gamma\equiv (q^2_c n_cA/lm_c)L_s$.
%is the dimensionless  parameter of the loop.
Combining the expression for the total flux,
\begin{eqnarray}
\label{current1}
f_t=f+\frac{L_sI}{\Phi_0},
\end{eqnarray}
with Eq. (\ref{pbc}), the wave vector is written as
\begin{eqnarray}
\label{kn} (1+\gamma)k_n l=2\pi n+2\pi f-\phi,
\end{eqnarray}
where $f\equiv\Phi_{\rm ext}/\Phi_0$.

%%%%%%%%%%%%%%%%%%%%%%%%%%%%%%%%%%%%%%%%%%%%%%%%%%%%%%%%%%%%%%%%%%%%%%%%%%%%%%%%%%%%%%%%%%%
%Figure 2
\begin{figure}[b]
\vspace*{25cm}
\hspace*{-9cm}
\includegraphics{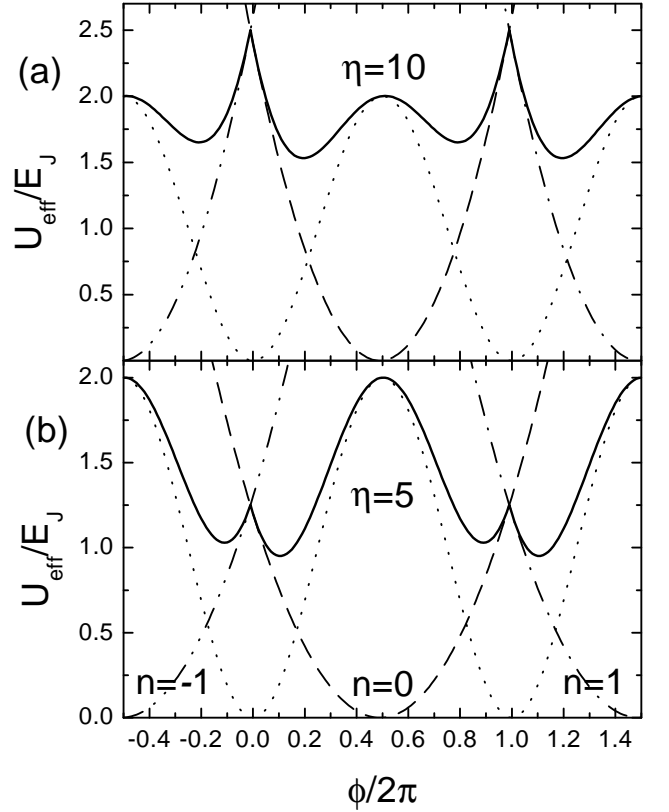}
\vspace*{-14cm}
\caption{Lowest
energy giving effective potential(solid line) in units of $E_J$
for $f$=0.49 as a function of Josephson phase for (a) $\eta$=10
and (b) $\eta$=5. The dotted  lines represent  the Josephson
coupling energy.}
\label{fig:WideWells}
\end{figure}
%%%%%%%%%%%%%%%%%%%%%%%%%%%%%%%%%%%%%%%%%%%%%%%%%%%%%%%%%%%%%%%%%%%%%%%%%%%%%%%%%%%%%%%%%%%

Neglecting the dissipative term originating from the quasi-particle
tunneling through the junction, Eq. (\ref{I})  becomes
\begin{eqnarray}
-(n_cAq_c/m_c)\hbar k_n=-I_c\sin\phi-C(\Phi_0/2\pi)\ddot{\phi}.
\end{eqnarray}
With the representation for $k_n$ in Eq. (\ref{kn}) we get
\begin{eqnarray}
\label{motion} \left(\frac{\Phi_0}{2\pi}\right)^2C\ddot{\phi}=
\frac{\epsilon_0}{(1+\gamma)\pi}\left(n+f-\frac{\phi}{2\pi}\right)-E_J\sin\phi,
\end{eqnarray}
where  $E_J=I_c\Phi_0/2\pi$ is the Josephson coupling energy and
$\epsilon_0\equiv 2\pi^2\hbar^2n_cA/m_cl$.
Eq. (\ref{motion}) describes the  motion of a particle with
kinetic energy $E_C=Q^2/2C$ with $Q=C(\Phi_0/2\pi)\dot{\phi}$
in an effective potential $U_{\rm eff}(\phi)$,
\begin{eqnarray}
U_{\rm eff}(\phi)=E_n+E_J(1-\cos\phi),
\end{eqnarray}
where
\begin{eqnarray}
\label{En}
E_n&=&\frac{\epsilon_0}{1+\gamma}\left(n+f-\frac{\phi}{2\pi}\right)^2.
\end{eqnarray}

The meaning of $E_n$ can be understood more clearly when
we  represent $E_n$ as $E_n=(1+\gamma)E_{\rm kin}$ using Eq. (\ref{kn}), where
$E_{\rm kin}=(\hbar^2k_n^2/2m_c) n_cAl$ is the kinetic energy of Cooper
pairs. Since the induced energy of the loop can be represented as
$E_{\rm ind}=(1/2)L_sI^2=\gamma E_{\rm kin}$, we can see that $E_n$ is the
sum  of the kinetic energy and the induced
energy \cite{Imry,Bloch}, i.e., $E_n=E_{\rm kin}+E_{\rm ind}$.
If we introduce the kinetic inductance $L_k\equiv m_cl/An_cq^2_c$,
the kinetic energy becomes $E_{\rm kin}=(1/2)L_k I^2$ and
the parameter $\epsilon_0$ in Eq. (\ref{motion}) is represented by $\epsilon_0=\Phi^2_0/2L_k$.
If we introduce the energy scale $E_K\equiv \Phi^2_0/2L_k=\epsilon_0$
and $E_L\equiv\Phi^2_0/2L_s$, the parameter $\gamma$ becomes
\begin{eqnarray}
\gamma=\frac{L_s}{L_k}=\frac{E_K}{E_L}
\end{eqnarray}
and the energy $E_n$ in Eq. (\ref{En}) can be rewritten as
\begin{eqnarray}
\label{EnNew}
E_n&=&\frac{E_K}{1+\gamma}\left(n+f-\frac{\phi}{2\pi}\right)^2.
%E_n&=&\frac{\Phi^2_0}{2(L_k+L_s)}\left(n+f-\frac{\phi}{2\pi}\right)^2.
\end{eqnarray}
The similar expression was obtained for a superconducting loop without Josephson junction
\cite{Majer}.
%Then the sum, $E_{\rm kin}+E_{\rm ind}$, can be simply expressed as $(1+\gamma)E_{\rm kin}$
%The energy $E_n$ can be considered as a sum of the kinetic energy of Cooper
%pairs, $E_{\rm kin}=(\hbar^2k_n^2/2m_c) n_cAl$, and the induced energy of the loop,
%$E_{\rm ind}=(1/2)L_sI^2$.
%The charging energy $E_C$ can be represented as $P^2/2M$ with
%$P=-i\hbar\partial /\partial \phi$  \cite{Mooij}
%in this new picture witch causes quantum tunneling between two wells in the
%effective potential $U_{\rm eff}(\phi)$.

The number of excess Cooper pair charges on Josephson junction
$\hat{N_c}\equiv \hat{Q}/q_c$ is conjugate to the phase difference $\hat{\phi}$
and  the commutation relation $[\hat{\phi},\hat{N_c}]=i$ gives the quantum phase fluctuations
of the junction.
The Hamiltonian $\hat{H}=\hat{P}^2/2M+U_{\rm eff}(\hat{\phi})$
describes the dynamics of the phase difference,
where $\hat{P}\equiv \hat{N_c}\hbar= -i\hbar\partial /\partial \hat{\phi}$
and $M=(\Phi_0/2\pi)^2C$.
The charging energy $E_C$ causes quantum tunneling between two wells in the
potential $U_{\rm eff}(\hat{\phi})$ of the rf-SQUID and  the following
three-Josephson junction loop.

The double-well structures shown in Fig. \ref{fig:WideWells} are given by taking the
lowest energy for the quantum states $n=-1,0,$ and $1$ for the
effective potential $U_{\rm eff}(\phi)$. The shape of the effective
potential is determined by the parameter $\eta$ defined by
\begin{eqnarray}
\eta\equiv\frac{E_K}{(1+\gamma)E_J}.
%=\frac{\Phi^2_0}{2(L_k+L_s)E_J}.
\end{eqnarray}
For large $\gamma$, the value of $\eta$ can be determined by two parameters
$L_s$ and $E_J$. We calculate the value of the  parameter $\gamma$ to obtain
$\gamma\approx 600$ for the parameters of  the rf-SQUID of Ref. [4].
%and $\gamma\approx 200$ for the three-Josephson junction loop of Ref.  [5].
At much low temperature T $\sim$ O(10mK),
we roughly estimate the Cooper pair density in a pure metal
$n_c\approx n/2$~ \cite{Tinkham}  with the conduction electron density $n$.
Even if we consider the effect reducing the Cooper pair density such as
band structure of the  materials, finite temperature, and so on,
we expect that the value of $\gamma$ is still large such that $\gamma>> 1$.
In this case the parameter $\eta$ is approximately represented as follows,
\begin{eqnarray}
\label{apprEta}
\eta\approx\frac{E_L}{E_J}=\frac{\Phi^2_0}{2L_s E_J}
=\frac{\pi\Phi_0}{L_s I_c},
\end{eqnarray}
where  $E_J=I_c\Phi_0/2\pi$ with critical current of the Josephson junction $I_c$.
We obtain  that $\eta\approx 8.5$ for the rf-SQUID \cite{Friedman}.
%and $\eta\approx 3300$ for the three-Josephson junction qubit \cite{Mooij}.

For the value of $\eta=10$, we can see the  double-well
structure in Fig. \ref {fig:WideWells} (a) with a smooth barrier centered at
$\phi/2\pi=0.5$ for $\Phi_{\rm ext}/\Phi_0$=0.49. The usual double-well structure with a smooth
barrier corresponds to this one. The left (right) well corresponds to the
diamagnetic  (paramagnetic) current state. Thus the total flux,
$\Phi_t=\Phi_{\rm ext}+\Phi_{\rm ind}$, for the state at left (right) well
approaches  $0 (\Phi_0)$ as the self inductance becomes large.
For a smaller $\eta$, however, a new double-well potential emerges
around the cusp as shown in Fig. \ref{fig:WideWells} (b). The high cusp barrier
shrinks while the low smooth barrier grows up as $\eta$ decreases. Thus
we get two wells separated by a cusp barrier.
The state of current at right (left) well is diamagnetic (paramagnetic).

At the local minima of $U_{\rm eff}(\phi)$, the condition
$\partial U_{\rm eff}(\phi)/\partial\phi=0$
gives $-(n_cA\hbar^2/m_c)k_n+E_J\sin\phi=0$ and
the current relation becomes
\begin{eqnarray}
\label{cr}
-\frac{n_cAq_c}{m_c}\hbar k_n+I_c\sin\phi=0.
\end{eqnarray}
Actually Eq. (\ref{current1}) is also a current relation, i.e.,
$I=(\Phi_t-\Phi_{\rm ext})/L_s$.
If one consider both $\Phi_t$ and
$I$ as independent dynamic variables while $\Phi_{\rm ext}$ is a control
variable and minimize the energy of the system with respect to the variable
$\Phi_t$, the current relation, Eq. (\ref{current1}), can be obtained \cite{Imry, Bloch}.
Now the remaining dynamic variables are $\phi$ and $I$ or equivalently $k_n$.
The current relation, Eq. (\ref{cr}), is obtained by  minimizing
the total energy with respect to the variable $\phi$.
This current relation warrants the validity of our effective potential.
%This current relation has not been tested in other studies
%using a superconducting ring with a Josephson junction.

\section{Three-Josephson junction loop}

Recently the superconducting loop with three-Josephson junctions
of Fig. \ref {fig:SingQubit} (b) has been proposed as an efficient qubit
\cite{Mooij}. The periodic boundary condition similar to Eq. (\ref{kn})
is written as
\begin{eqnarray}
\label{pbc3} (1+\gamma)k_n l=2\pi n+2\pi f-\phi_1-\phi_2-\phi_3.
\end{eqnarray}
The effective potential $U_{\rm eff}(\phi_1,\phi_2,\phi_3)$
for this qubit is given by
\begin{eqnarray}
U_{\rm eff}(\phi_1,\phi_2,\phi_3)&=&E_n+E_J(1-\cos\phi_1)+E_J(1-\cos\phi_2)\nonumber\\
&+&E_{J3}(1-\cos\phi_3),
\end{eqnarray}
where
\begin{eqnarray}
E_n&=&\frac{E_K}{1+\gamma}
\left(n+f-\frac{\phi_1+\phi_2+\phi_3}{2\pi}\right)^2.
\end{eqnarray}
Here we introduced the rotated coordinates
$(\phi_a,\phi_b,\phi_c)$ such that $\phi_a\equiv
(1/2)(\phi_1+\phi_2), \phi_b\equiv (-1/2)(\phi_1-\phi_2),$ and
$\phi_c\equiv (1/3)(\phi_1+\phi_2+\phi_3)$.
Then the effective potential of this qubit is written as,
\begin{eqnarray}
\label{3JJUeff}
&&U_{\rm eff}=\frac{E_K}{1+\gamma}
\left(n+f-\frac3{2\pi}\phi_c\right)^2\nonumber\\
&&+2E_J(1-\cos\phi_a\cos\phi_b)
+E_{J3}\left[1-\cos(3\phi_c-2\phi_a)\right].\nonumber\\
\end{eqnarray}

%%%%%%%%%%%%%%%%%%%%%%%%%%%%%%%%%%%%%%%%%%%%%%%%%%%%%%%%%%%%%%%%%%%%%%%%%%%%%%%%%%%%%%%%%%%
%Figure 3
\begin{figure}[t]
\vspace*{8.5cm}
\includegraphics{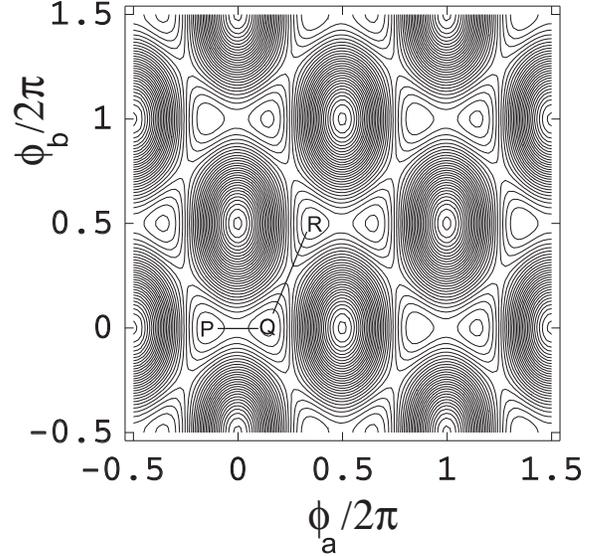}
\vspace{-0.5cm}
\caption{The effective potential of three-Josephson junction loop
in ($\phi_a,\phi_b$) plane, where $E_{J3}/E_J$=0.8,
$f$=0.495, and $\eta$=1000. We set $\phi_c/2\pi$=$f$/3. }
\label{fig:PhiaPhib}
\end{figure}
%%%%%%%%%%%%%%%%%%%%%%%%%%%%%%%%%%%%%%%%%%%%%%%%%%%%%%%%%%%%%%%%%%%%%%%%%%%%%%%%%%%%%%%%%%%

%%%%%%%%%%%%%%%%%%%%%%%%%%%%%%%%%%%%%%%%%%%%%%%%%%%%%%%%%%%%%%%%%%%%%%%%%%%%%%%%%%%%%%%%%%%
%Figure 4
\begin{figure}[t]
\vspace*{7.5cm}
\includegraphics{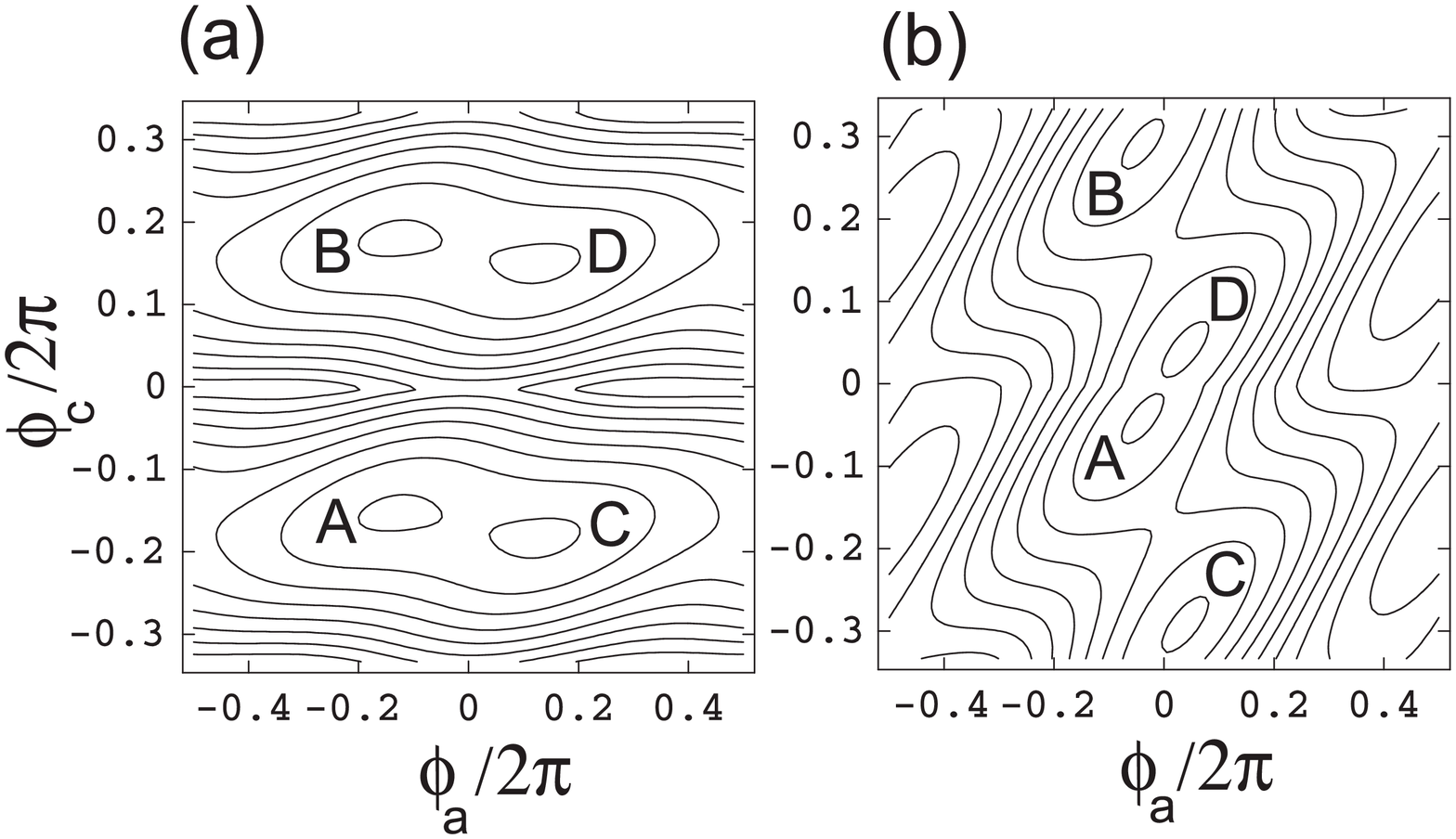}
\vspace*{5.5cm}
\includegraphics{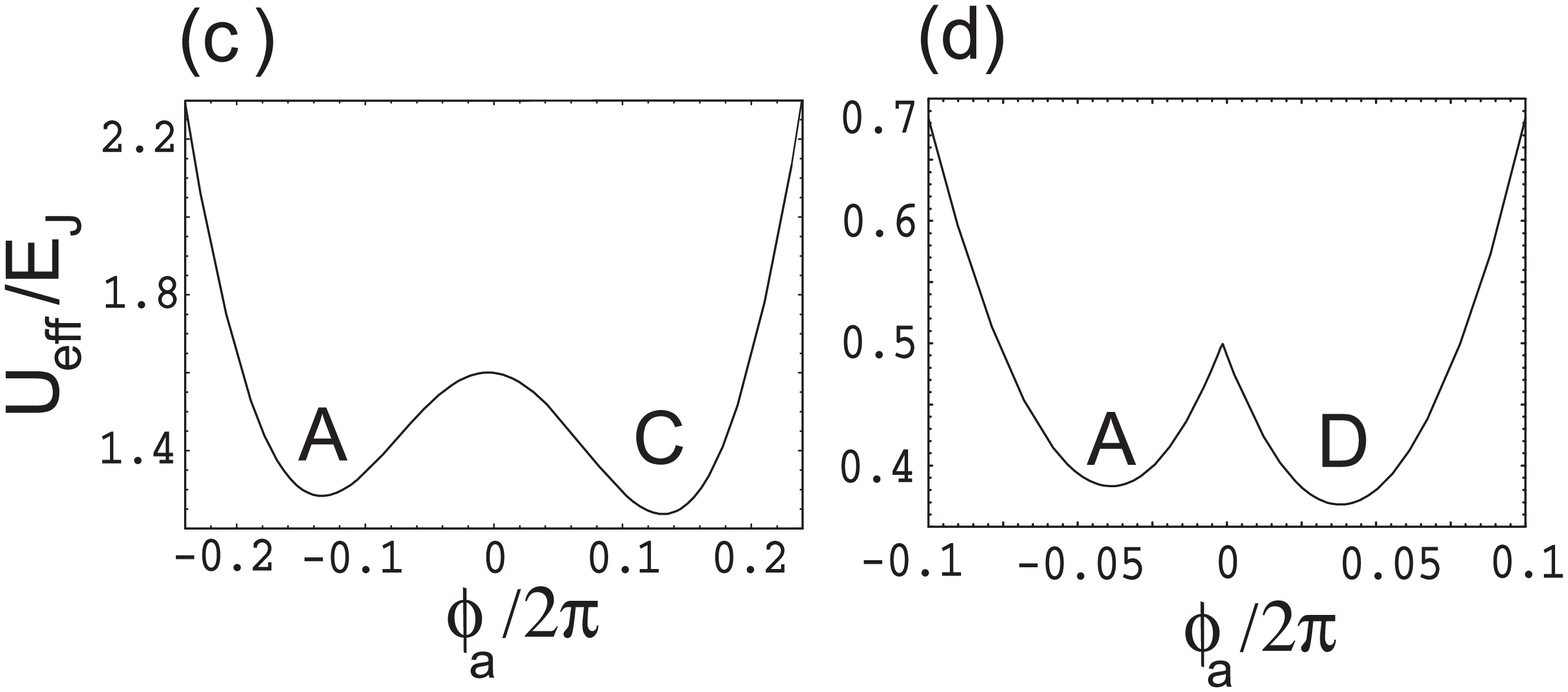}
\vspace{-3cm}
\caption{ Contour plots for the effective
potential of three-Josephson junction qubit with (a)$\eta$=50 and
(b)$\eta$=2 when $f$=0.495 and $\phi_b$=0. ${\rm A, B, C}$, and
${\rm D}$ denote the ground states at the local potential minima.
The lines $\phi_c\approx 0$ in (a) and (b) represent the cusp barrier.
The potential in (c) is drawn along the line connecting the points
${\rm A}$ and ${\rm C}$ for $\eta=50$, while the potential in (d)
is for the points ${\rm A}$ and ${\rm D}$ for $\eta=2$.}
\label{fig:3JJ}
\end{figure}
%%%%%%%%%%%%%%%%%%%%%%%%%%%%%%%%%%%%%%%%%%%%%%%%%%%%%%%%%%%%%%%%%%%%%%%%%%%%%%%%%%%%%%%%%%%

Fig. \ref{fig:PhiaPhib} shows the effective potential at
$\phi_c/2\pi=f/3$, where we can see that the value of $\phi_b$
is strongly localized around $\phi_b=n\pi$ since the barrier
between $Q$ and $R$ is much higher than that between $P$ and $Q$.
This means that the phase differences at the potential minima
in  identical Josephson junctions are $2\pi n$.
For $\phi_b=0$ the effective potential is shown in Fig. \ref{fig:3JJ}.
The barriers between $A$ and $C$ as well as between $B$ and $D$ are smooth, as
can be seen in Fig. \ref{fig:3JJ} (c). The two current states at $A$ and $C$
correspond to the qubit states of Ref. [5], where $\eta$ is so large that the cusp
barrier is high enough for tunneling across the barrier to be neglected.
The cusp barrier between $A$ and $D$ shrinks while the smooth barrier
grows as  $\eta$ decreases. In Fig. \ref{fig:3JJ} (d), thus,  we can see
the cusp barrier between $A$ and $D$.
The tunneling through this cusp barrier becomes dominant as $\eta$ becomes small.

The current in the loop is given by $I=-I_c\sin\phi_a$. Thus the
current states at $A$ and $B$ ($C$ and $D$) are in a paramagnetic
(diamagnetic) current state. The values of $\phi_a$ and $\phi_c$
at potential minima can be obtained through the current relations,
$\partial U_{\rm eff}/\partial \phi_i=0$ which give
\begin{eqnarray}
\frac{E_K}{(1+\gamma)\pi}\left(n+f-\frac{\phi_1+\phi_2+\phi_3}{2\pi}\right)
=E_{Ji}\sin{\phi_i},%\nonumber\\
\end{eqnarray}
where $E_{J1}=E_{J2}=E_J$.
In the rotated coordinate we obtain the relations
\begin{eqnarray}
\label{cr3JJ}
\frac{\eta}{\pi}\left[n+f-\frac{3}{2\pi}\phi_c\right]&=&\sin\phi_a\nonumber\\
&=&\lambda\sin(3\phi_c-2\phi_a),
\end{eqnarray}
where $\lambda\equiv E_{J3}/E_J$ and we set $\phi_b=0$.
From  Eqs. (\ref{pbc3}) and (\ref{cr3JJ}), we calculate
$\tilde{f}\equiv (1+\gamma)k_n l/2\pi$.
We calculate the values of the  parameter $\gamma$ to obtain
$\gamma\approx 200$ and  $\eta\approx 3300$
for the three-Josephson junction qubit of Ref. [5].
For $\gamma >> 1$, the induced flux  in Eq. (\ref{ft}) can be represented such that
$\Phi_{\rm \rm ind}/\Phi_{0}=-\gamma k_n l /2\pi\approx -\tilde{f}$.

\begin{figure}[b]
\vspace*{7cm} \includegraphics{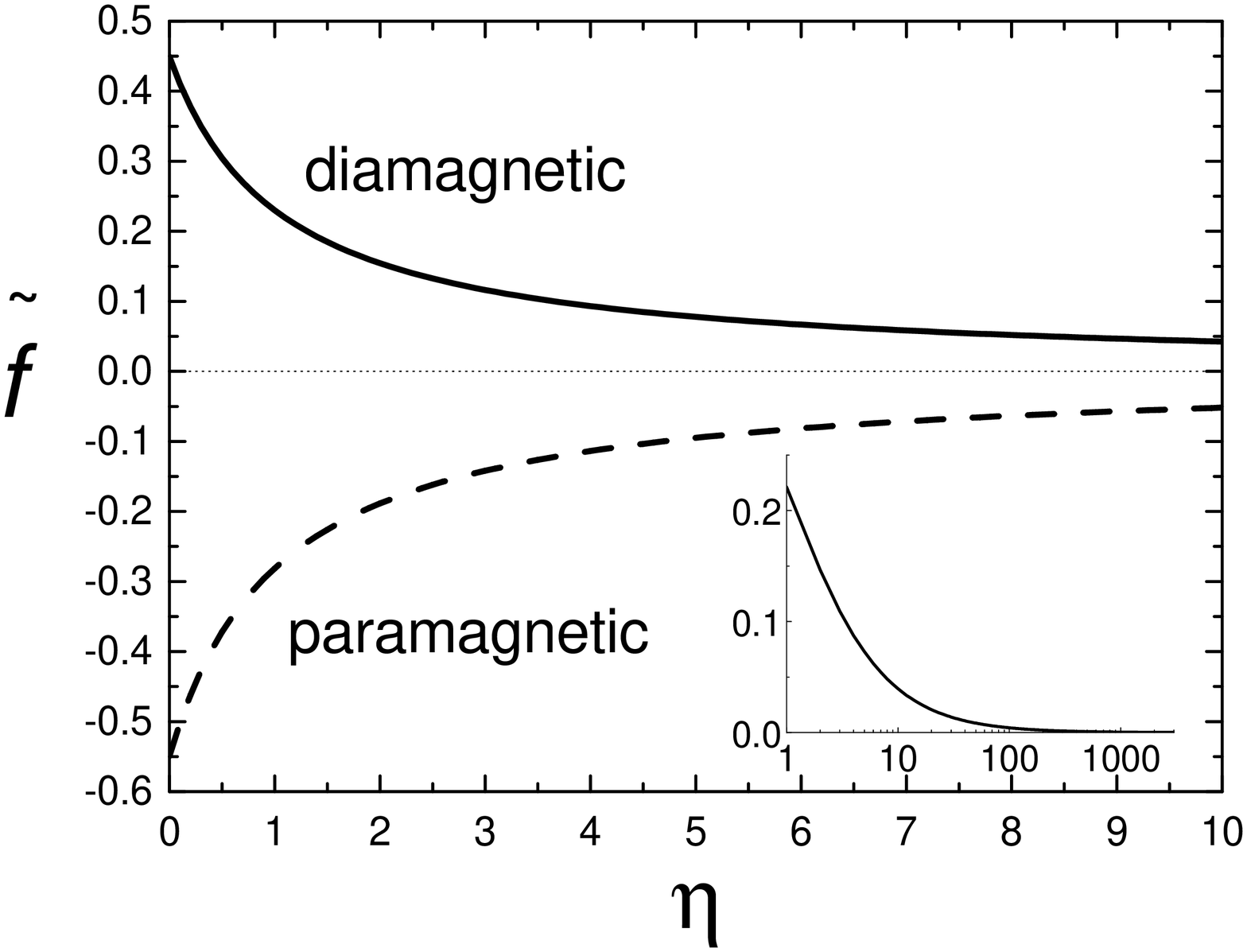}
\vspace{0cm}
\caption{Induced diamagnetic (solid line) and paramagnetic (dashed line)
fluxes as a function of $\eta$ when
$f$=0.45. Inset shows the induced diamagnetic flux for large
$\eta$.} \label{fig:indflux}
\end{figure}

In Fig. \ref{fig:indflux} the induced flux is shown when $f=0.45$ and
$\lambda=0.8$. The ground states in $C$ and $D$ induce a
diamagnetic current while the ground states in $A$ and $B$ induce
a paramagnetic current.
In the inset of Fig. \ref{fig:indflux} we can see that $\tilde{f}$ is very small for the qubit
of Ref.  [5] so that the induced flux is negligible and
the boundary condition, Eq. (\ref{pbc3}), becomes $2\pi n+2\pi f=(\phi_1+\phi_2+\phi_3)$
which is used in Ref.  [5].
On the contrary, when the value of $\eta$ is much smaller, the induced
flux become larger.  The value of  total flux
approaches 0 ($\Phi_0$) for diamagnetic (paramagnetic) current,
which clearly shows the flux quantization condition.
Thus we can see that the total flux becomes quantized as $L_s E_J$ increases,
while the fluxoid quantization condition,  Eq. (\ref{fluxoid}), is always satisfied.
For superconducting ring without Josephson junction, the total flux is quantized
when $L_s$ is sufficiently large \cite{Imry}.

%%%%%%%%%%%%%%%%%%%%%%%%%%%%%%%%%%%%%%%%%%%%%%%%%%%%%%%%%%%%%%%%%%%%%%%%%%%%%%%%%%%%%%%%%%%
%Figure 5
%%%%%%%%%%%%%%%%%%%%%%%%%%%%%%%%%%%%%%%%%%%%%%%%%%%%%%%%%%%%%%%%%%%%%%%%%%%%%%%%%%%%%%%%%%%

%%%%%%%%%%%%%%%%%%%%%%%%%%%%%%%%%%%%%%%%%%%%%%%%%%%%%%%%%%%%%%%%%%%%%%%%%%%%%%%%%%%%%%%%%%%
%Figure 6
\begin{figure}[t]
\vspace*{8.5cm}
\includegraphics{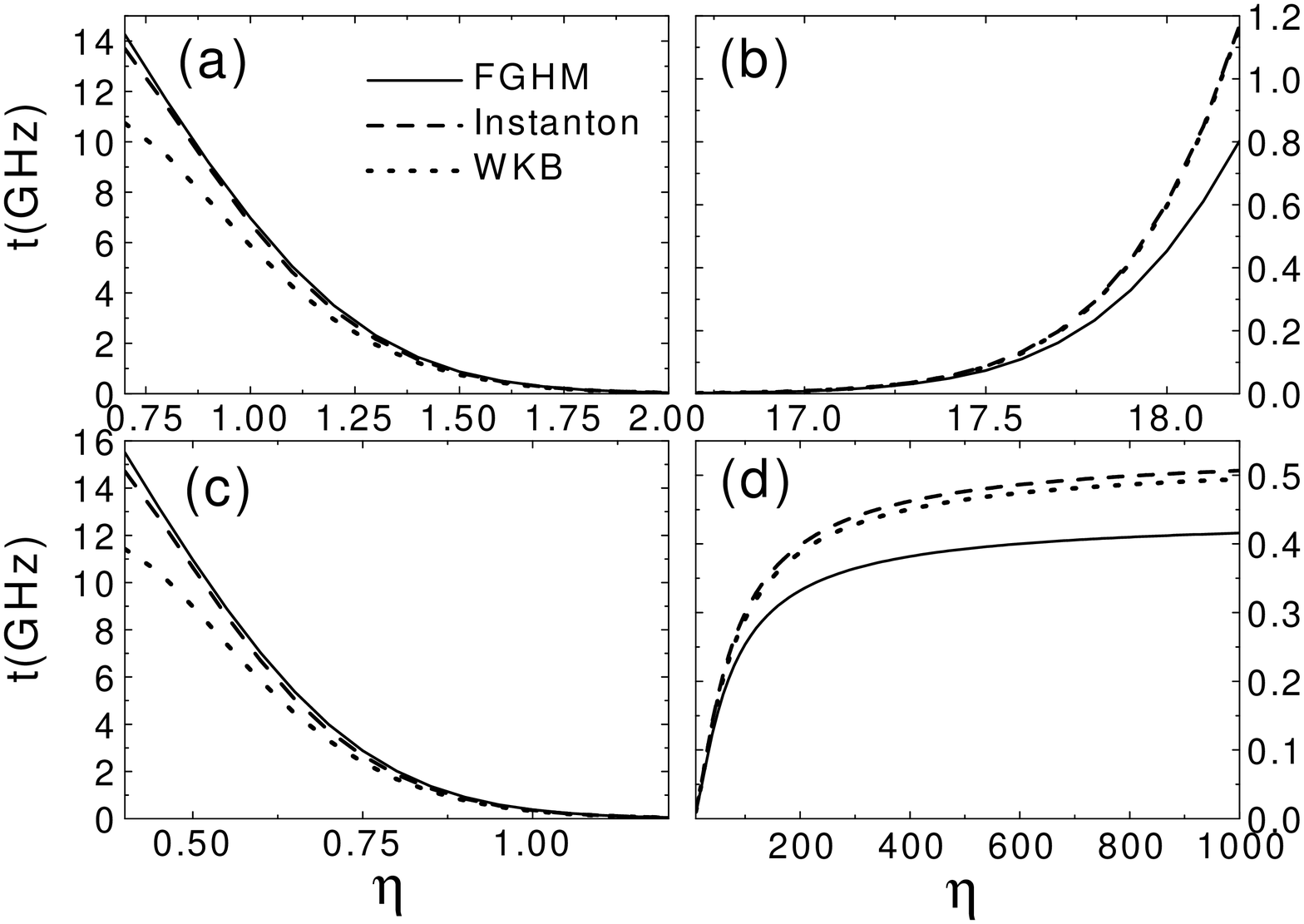}
\vspace*{-0.cm}
\caption{Tunneling amplitude for (a) cusp barrier and (b) smooth barrier
 in the effective potential of rf-SQUID, where we set $\alpha$=64000 (8000)
 and $E_J$=300K (75K) for the cusp (smooth) barrier.
 Figures (c) and (d) are for the three-Josephson junction qubit, where
 we set $\alpha$=64000 (80) and $E_J$=300K (10K) for the cusp (smooth) barrier.}
\label{fig:Tunn}
\end{figure}
%%%%%%%%%%%%%%%%%%%%%%%%%%%%%%%%%%%%%%%%%%%%%%%%%%%%%%%%%%%%%%%%%%%%%%%%%%%%%%%%%%%%%%%%%%%

\section{Tunneling frequency}

The quantum states in a cusp potential may be used to make a
more efficient qubit because of high tunneling frequency  across the
barrier. This allows more single qubit operations in a short dephasing time.
The geometric configuration of the cusp potential,
such as its higher curvature and smaller width, causes higher tunneling frequency.
We compare the tunneling rate calculated by the WKB approximation,
the instanton method \cite{Weiss}, and the Fourier Grid Hamiltonian
Method(FGHM) \cite{FGHM} in Fig. \ref{fig:Tunn}. The instanton method calculates
the tunneling rate $t$  as follows,
\begin{eqnarray}
t=\hbar\omega_0 C_0\sqrt{\frac{S_{\rm inst}}{2\pi\hbar}}
\exp\left[-\frac{S_{\rm inst}}{\hbar}\right],
\end{eqnarray}
where $S_{\rm inst}=M\int_{-\infty}^{\infty}d\tau\dot{\phi}^2(\tau)$,
$M=(\Phi_0/2\pi)^2C=\hbar^2/8E_c$, $\omega_0=\sqrt{k/M}$ is the
frequency of small oscillation in a well  with $k=\partial
^2U_{\rm eff}(\phi)/\partial\phi ^2|_{\phi_{\rm min}}$, and $C_0$
is a factor which depends on the shape of the double-well potential.
The instanton action is evaluated in such a way that $S_{\rm
inst}=\sqrt{2M}\int_{-\phi_{\rm min}}^{\phi_{\rm min}}
\sqrt{U_{\rm eff}(\phi)} d\phi$, where $-\phi_{\rm min}$ and
$\phi_{\rm min}$ represent  two potential minima. The instanton
satisfies the equation of motion $(1/2)M\dot{\phi}^2(\tau)=U_{\rm
eff}[\phi(\tau)]$. Integrating the equation of motion to obtain
the instanton trajectory and comparing it with the asymptotic
behavior, $C_0$ is given by
\begin{eqnarray}
C_0=\phi_{\rm min}\sqrt{\frac{2M\omega_0}{S_{\rm inst}}}\exp[\omega_0 I],
\end{eqnarray}
where $I$ is a constant obtained numerically.

The values of $E_C$ are chosen through the parameter $\alpha\equiv
E_J/E_C$ which is roughly proportional to  $1/\eta$ in experiments
\cite{Mooij,Wal,Friedman}. In Fig. \ref{fig:Tunn} we see that the
frequency of tunneling  through the cusp barrier for low $\eta$ is
much higher  than that for the smooth barrier in both rf-SQUID and
the three-Josephson junction qubit.
The tunneling through the potential barriers in the  effective potential,
$U_{{\rm eff}}$, introduces the tunneling term in the single qubit Hamiltonian,
\begin{eqnarray}
\label{Hqubit}
H_{\rm qubit}=F(\Phi_{\rm ext})\sigma^{z}-t\sigma^x,
\end{eqnarray}
where $F(\Phi_{\rm ext})$ is the energy level separation between two current states
and $\sigma^{z (x)}$ are the Pauli matrices.
The first (second) term in Eq. (\ref{Hqubit}) causes the single qubit rotation, $R_z (R_x)$.
The operation time $\tau_{\rm op}$ for the $R_x$ gate
is inversely proportional to the tunneling frequency.
Since the quality factor is $Q_{\phi}=\tau_{\phi}/\tau_{\rm op}$
with the dephasing time $\tau_{\phi}$,
%and the operation time $\tau_{\rm op}$,
the present results show that the cusped
barrier allows the quality factor that is an order of magnitude
larger than occur with the smooth barrier.

Lower $\eta$ requires higher  $E_J$ as shown in Eq. (\ref{apprEta}).
Thus,  a higher inductive flux in Fig. \ref{fig:indflux} could be detrimental
to the phase coherence of the Cooper pair wave function.
However, recent experiments for a three-Josephson junction
loop with a Cooper-pair island \cite{Vion} and  a current-biased Josephson junction
qubit \cite{Martinis} show that using a large Josephson junction
the high quality factor can be achieved.
A current-biased qubit of dc SQUID type
%with two Josephson junctions may
%consist of a loop with a piercing magnetic flux and two leads. This qubit
with large Josephson junctions can be described
by the present theory for low value of $\eta$.
%and, thus, a experimental effort is invoked.

%The qubit using the cusp potential for low $\eta$ is much less
%influenced by the quantum phase slip.
The quantum phase slip  may be relevant in a thin superconducting loop
when the energy levels are degenerate \cite{Matveev}.
%In Fig. \ref{wideWells} the cusp corresponds to the degenerate point
%of different values $n$ in Eq. (\ref{EnNew}).
The phase slip process causes the change of tunneling frequency
and, thus, induces the decoherence.
For the lower  value of $\eta$ in Eq. (\ref{apprEta})
the loop inductance is larger, which means the loop is  thicker.
Therefore the qubit using the cusp barrier with the lower value of $\eta$
as shown in  Fig. \ref{fig:WideWells} may have  a longer dephasing time
than that using the smooth one.

\section{Summary}

Using the boundary condition for the phase evolution of
the wave function of the Cooper pairs,
we obtain a new double-well  potential with cusp
barriers for both rf-SQUID and  three-Josephson junction loop.
At the minimum of the effective potential we show that
the current relation is satisfied.
This picture allow us to calculate the wave vector of Cooper pairs, $k_n$,
as a function of $\eta$ and, thus, obtain the induced flux
to show that the flux quantization condition is satisfied when
the value of $\eta$ becomes small.
Moreover we calculate the tunneling
frequency as a function of $\eta$ which is the characteristic
parameter responsible for the shape of the potential.
The tunneling frequency in the cusp potential is much higher
than that in the smooth potential, thus,
the cusp potential offers a more efficient qubit.

\begin{center}
{\bf ACKNOWLEDGMENTS}
\end{center}

This work is supported  by the Korean Science and Engineering Foundation
through the Center for Strongly Correlated Materials Research (2002)
at Seoul National University and the Brain Korea 21 Project in 2002.

\end{document}